\documentclass[fleqn,10pt]{wlscirep}
\usepackage[utf8]{inputenc}
\usepackage[T1]{fontenc}
\usepackage{graphicx}
\usepackage{caption}
\usepackage{subcaption}
\usepackage{float}
\usepackage[mathscr]{euscript}

 \newcommand{\sw}[1]{{\color{blue}{ #1}}}

\title{A new architecture for high speed core-selective switch for multicore fibers}

\author[1]{Cristóbal~Melo}
\author[2]{Matías~Reyes.~F.}
\author[1]{Diego Arroyo}
\author[3]{Esteban~S.~Gómez }
\author[3,4]{Stephen~P.~Walborn}
\author[3,4]{Gustavo~Lima}
\author[1]{Miguel~Figueroa}
\author[2]{Jaime~Cariñe}
\author[1,*]{Gabriel~Saavedra}

\affil[1]{Department of Electrical Engineering, Universidad de Concepción, Concepción, Chile}
\affil[2]{Department of Electrical Engineering, Universidad Católica de la Santísima Concepción, Concepción, Chile}
\affil[3]{Department of Physics, Universidad de Concepción, Concepción, Chile}
\affil[4]{Millennium Institute for Research in Optics, Universidad de Concepción, 160-C, Concepci\'on, Chile}
\affil[*]{gasaavedra@udec.cl}

\begin{abstract}
 The use of multicore optical fibers is now recognized as one of the most promising methods to implement the space-division multiplexing techniques required to overcome the impending  capacity limit of conventional single-mode optical fibers. Nonetheless, new devices for networking operations that are compatible with these fibers will be required in order to implement the next-generation of high-capacity  optical networks. In this work, we develop a new architecture to build a high-speed core-selective switch, which is critical for efficiently distributing signals over the network. The device relies on multicore interference, and can change among outputs in less than 0.7 $\mu$s, while achieving less than -18 dB of average inter-core crosstalk, making it compatible with a wide range of network switching tasks. The functionality of the device was demonstrated by routing an optical signal modulated at 1GBs and also by successfully switching  signals over a field-installed multicore fiber network.  Our results demonstrate for the first time the operation of a multicore optical fiber switch functioning under real-world conditions, with switching speeds that are three orders of magnitude faster than current commercial devices. This new optical switch design is also fully compatible with standard multiplexing techniques and, thus, represents an important achievement towards the integration of high-capacity multicore telecommunication networks.
\end{abstract}
\begin{document}

\flushbottom
\maketitle

\thispagestyle{empty}

\section*{Main}
Optical fibers are essential for modern high-speed communication systems, with an ubiquitous presence in access, metropolitan and inter-continental networks. To satisfy the increasing demand for data transmission, optical fibers need to increase their information carrying capacity, and to do so, space division multiplexing (SDM) has been acknowledged as one of the potential solutions to the impending capacity crunch of optical fiber links and networks\cite{richardson10,richardson13,Puttnam:21,Winzer:17}. The main idea behind SDM is to employ specially designed fibers that can support multiple optical spatial modes, thus providing new transmission channels within the fiber that can increase the multiplexing capabilities. Nonetheless, several challenges still need to be resolved for the deployment of high-capacity SDM optical networks \cite{Puttnam:21,ON_mendinueta}. For instance,  the development of devices such as optical amplifiers, multiplexers and optical switches is crucial for their integration with the actual telecommunication infrastructure.

Among the candidate technologies for SDM, multicore fibers (MCF) - fibers with multiple cores in a single cladding with standard fiber diameter - have shown potential to transmit large data rates for short- and long-reach links. In this case, each available core mode is used as an extra and independent transmission channel, each one being compatible with standard wave-division multiplexing (WDM) techniques.  Recently, throughputs of 10.66 Pbps over a distance of 13 km\cite{10.66Ptbit}, and 10.16 Pbps over 11km\cite{Soma:10.16-Peta-B/s}, were demonstrated  using  few-mode multicore fibers (FM-MCF). For long reach systems, transmission of up to 120 $\times$ 100 Gbps channels in 4-core MCFs was performed over 2,768, 4,014, and 5,350 km \cite{Ryf:2768-4014km,Takeshita:5350km}, while using 7-core MCFs, as many as 201 $\times$ 100 Gbps channels were transmitted over a transatlantic link of 7,326 km\cite{Morita:14}.

Despite the aforementioned breakthroughs, operating an optical network requires the network manager to successfully allocate resources for each of the users. To do so, the selection of a core to transmit a given signal and devices capable of performing this operation will be a critical part of operating an MCF optical network. Depending on the property used to switch a signal in an optical node, four switching granularities have been identified as requisites for multi-dimensional SDM nodes\cite{Marom}, namely: independent spatial mode/wavelength channel switching; spatial mode switching across all wavelength channels; wavelength switching across all spatial modes; and, wavelength switching across spatial mode subgroups.   

For MCFs, the switching of a spatial mode is reduced to the selection of a core within the fiber where the information will be transmitted, usually termed core-switching.  All-fiber solutions to perform core-switching have been reported using long period gratings (LPG)\cite{LPG}, where the switching of signals between cores was performed by bending a MCF inscribed with LPG.
With this switching technique, transmission of 6 $\times$ 224 Gbps optical channels was achieved, with a minimum insertion loss of 11.3 dB and a maximum extinction ratio of 39 dB\cite{LPG}. Another approach to perform core-switching in MCF is beam steering  \cite{BeamSteering,BS_exp_pot}. These systems are based on a folded telecentric lens configuration that can steer the optical signal from the output core of a MCF into a different input core of a second MCF. In a recent result \cite{BS_exp_pot}, switch losses between 1.23 dB to 2.21 dB and crosstalk between cores lower than -34 dB were reported. Mechanical core switches have also been proposed, where a single motor is used to rotate a MCF to align the cores with a single-mode fiber, thus allowing the selection of a particular transmission core in the MCF \cite{RotatedMCF}. As the transmission core was varied, the connection loss ranged from 0.39 dB to 1.30 dB, and the reported crosstalk was as low as -60 dB. For the switches mentioned above, switching speeds on the order of milliseconds were achieved. 
However, to make MCFs compatible with the high-speed switching that is essential for modern optical communication applications, such as data center networking, or for datacom applications, switching speeds below 10~$\mu$s are desired \cite{Yoo}.

Here we report a fast, all-fiber core-selective switch to use with MCFs. It relies on a GHz rate re-configurable interferometric array for core-switching. The MCF switch can route an optical signal to a new output on a timescale of about 0.7$\mu$s, making it three orders of magnitude faster than previous MCF switches.  Favorable characteristics such as reasonable insertion loss (7.7dB), high extinction ratio (19.8dB) and low inter-core cross talk ($-16.25$ dB average) make this a promising approach for future MCF-based optical networks.  To showcase the functionality, we use the MCF switch to route a 1GBs optical signal among the four cores of a MCF links, obtaining bit error rates of $10^{-9}$. To fully demonstrate the compatibility of our switch with future MCF networks, we tested it by routing an optical signal around a field-installed MCF network.  To our knowledge, this is the first time a a signal is directed through a MCF network with high switching speeds.

\section*{Results}

\begin{figure}[b!]
    \centering
    \includegraphics[width=\textwidth]{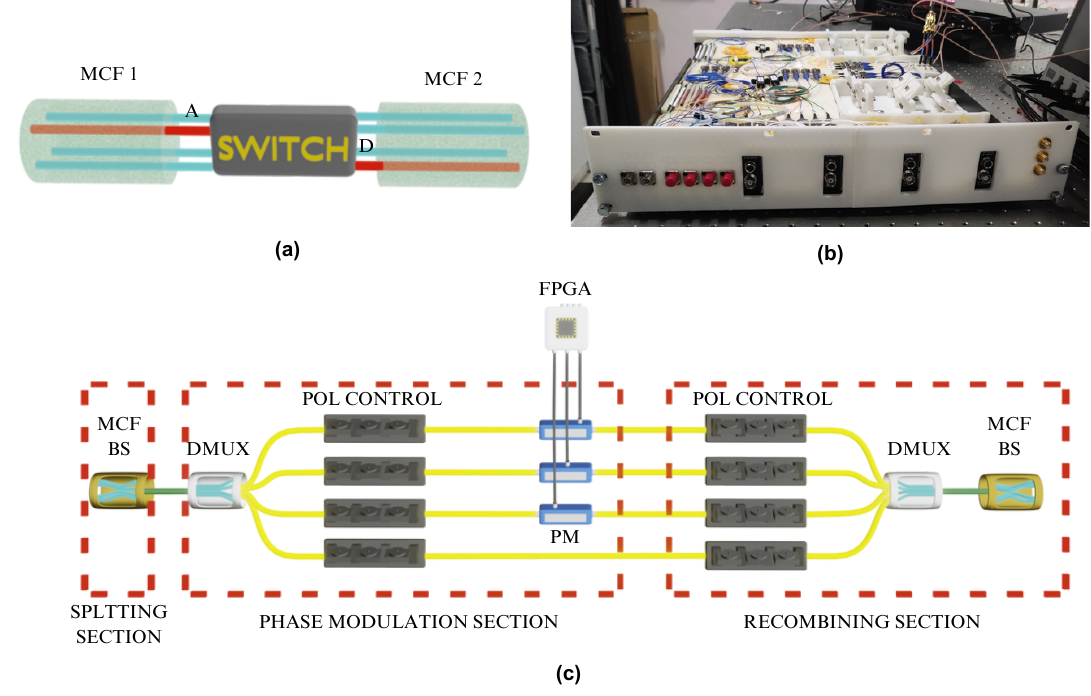} 
    \caption{\textbf{Implementation and Operation of the Multicore Fiber Switch.} \textbf{(a)}, Switch concept between 2 multicore fibers. \textbf{(b)}, Photo of MCF switch implemented in the laboratory. \textbf{(c)}, Conceptual design of the MCF switch.}
    \label{fig:EX_OP}
\end{figure}

\subsection*{Design of the MCF core-selective switch}
The main function of a core-selective switch is to control the path of optical signals in the spatial domain. That is, to route all the wavelengths present in--say-- core A in fiber 1 into core D of fiber 2, as illustrated in Fig.~\ref{fig:EX_OP}\textcolor{blue}{a}. The MCF switch was constructed into a standard 19'' rack shelf compatible with telecommunications infrastructure, as shown in Fig.~\ref{fig:EX_OP}\textcolor{blue}{b}. The conceptual design of the switch architecture is illustrated in Fig.~\ref{fig:EX_OP}\textcolor{blue}{c}.  The operation of the device can be  
can be divided into splitting, phase modulation, and recombination sections.  Additionally, a digital control system is responsible for performance and output selection. Let us briefly describe the operation of the device:

\textbf{\textit{Splitting section: }}  The input MCF fiber is connected to a multicore fiber beam splitter (MCF-BS), which splits the optical signals entering the switch with a split ratio of  25\% \cite{Carine:20}.

\textbf{\textit{Phase modulation section: }} Following the splitting section, a commercial MCF-SMF de-multiplexer (DMUX) was used to separate the MCF-BS into four independent single mode fiber (SMF) paths, all but one of which are connected to Lithium Niobate Electro-Optic phase modulators (PM). Each PM contains an internal polarizer, thus a polarization controller (POL CONTROL) was included before each PM.
The PMs were connected to a field programmable gate array (FPGA) via an amplifier circuit, allowing the FPGA to perform controlled phase shifts over three of the four fibers.

\textbf{\textit{Recombination section:}} Here a second MCF-BS is used to recombine the four processed signals. Upon recombination, optical interference occurs, such that the output intensity in each fiber core is dependent upon the phase shifts applied by the three PMs. This is the physical process behind the operation of the switch. For this to occur perfectly,  each path must have the same optical path length, the same optical losses, and introduce the same polarization variations to maximize the interference effect. For this reason, a fiber delay line in each path to equalize the length of the paths and a second set of polarization controllers was used. 
Note that all fiber components were connected using commercial FC connectors and adapters. More details are provided in the Methods section. 

As mentioned above, a digital control system was used to stabilize the phase fluctuations within the paths of the interferometer, and to select the desired routing operation. 
The digital system is composed of an FPGA, connected to the PMs through four digital-analog converters (DACs) and an amplifier circuit. At the output of the switch, the output power in each core was monitored and connected to the FPGA to apply a phase stabilization algorithm\cite{Carine:20}. Additionally, the FPGA was able to select a set of phases to perform the core switching process. 

Stable device operation is shown in Fig.~\ref{fig:phase_stabilization}, where from 0 to 7 s, no phase stabilization is used and the optical power is seen to randomly couple between all cores. While the phase stabilization algorithm is operating, the optical signal is seen to exit the device only from core 1.
\par
With the interferometer stabilized, the insertion loss of the device was measured. An average insertion loss of 7.7 dB was observed when entering the device from core 1. The main causes of the insertion loss are the use commercial Lithium Niobate Modulators that present average losses of 3.3 dB, and the combination of DMUX, MCF-BS and DMUX used in the splitting and recombining sections present average losses of 2.2 dB, respectively. The measured insertion loss can be compared to the ones obtained in all-fiber devices \cite{LPG}, however free-space switch implementations \cite{BS_exp_pot} or micro-electromechanical systems (MEMS) can offer lower insertion losses.

\subsection*{Core switching}
\label{CoreSwitching}

\begin{figure}[t!]
  \centering
  \centering
    \begin{subfigure}[b]{0.33\textwidth}
        \includegraphics[width=\textwidth]{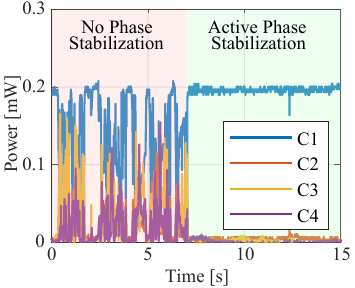}
        \caption{}
        \label{fig:phase_stabilization}
    \end{subfigure}
    \hfill
  \begin{subfigure}[b]{0.33\textwidth}
    \includegraphics[width=\textwidth]{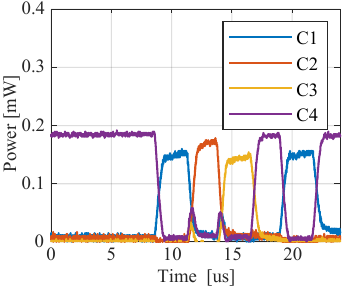} 
    \caption{}
    \label{fig:SW_2.5}
  \end{subfigure}%
  \begin{subfigure}[b]{0.33\textwidth}
    \includegraphics[width=\textwidth]
     {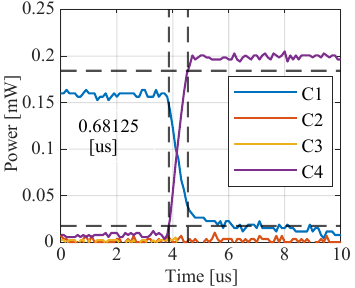}
    \caption{}
    \label{fig:RiseTime}
  \end{subfigure}
  \caption{\textbf{Output power for each core as a function of time.} \textbf{(a)}, Output power for every core with: no phase stabilization (0-7 s), active phase stabilization (7-15 s). \textbf{(b)}, Core switching at 2.5 $\mu$s. \textbf{(c)}, Rise time of the switching operation.}
  \label{fig:Coreswitching}
\end{figure}
In order for the device to operate as an optical switch, it needs to be able to select one of the output cores. To do this, a set of phases is applied to each PM to set the desired output core. Commercial PMs\cite{ThorlabsLN65S} with operating bandwidth on the order of GHz were used, while the FPGA driver system used a DAC121S101 converter with a conversion rate of 0.8 MSps considering RF filters. With this, the proposed device can operate on timescales compatible with standard network physical layer switching speeds ($\sim$100~$\mu$s)\cite{Yoo} and flow-switching or burst-switching speeds($\sim$ 25~$\mu$s)\cite{Zhang}.

To demonstrate operation of the MCF switch, we connected a continuous wave (CW) laser  (at 1550 nm) to the system. With the digital control system in operation, we configured the FPGA to switch the output core every 2 samples, allowing to switch the signal every 2.5~$\mu$s as shown in Fig.~\ref{fig:SW_2.5}. Stable switching performance is observed, as the light is able to be redirected to all of the 4 output cores. 
Fig.~\ref{fig:RiseTime} shows the observed rise time when the device is performing the core-switching operation. A rise time of less than 0.7~$\mu$s was achieved, allowing 1.8 ~$\mu$s of operation for data transmission between a change in transmission core, making the MCF switch compatible with burst-switching applications.
Considering the measured rise time, we can see that the proposed system can operate at switching speeds three orders of magnitude faster than previous solutions for MCF switching \cite{LPG,BS_exp_pot,RotatedMCF}.
\par
 In addition to low insertion loss and fast switching speed, the MCF switch displayed favorable signal extinction and cross-talk characteristics. An extinction ratio of 19.8 dB, which was calculated by comparing the signal in output core 1 at the high and low power regions in Fig.~\ref{fig:SW_2.5}. The  inter-core cross-talk (IC-XT) was calculated as the ratio between the powers in the output cores using $IC-XT=10log_{10}(P_{out,i}/P_{out,j)}$ dB, where $P_{out,i}$ is the selected output core $i$ and $P_{out,j}$ is the adjacent core $j$\cite{crosstalkdef}. An average IC-XT value of -16.25 dB was observed, with a minimum of -22 dB and a maximum of -12 dB. 
Finally, a second tuneable CW laser was used to evaluate WDM compatibility. The device was stabilized using the laser centered at 1550 nm, while the visibility of the interferometer was evaluated for wavelengths between 1527 and 1569~nm. Further details of WDM operation are found in section \textbf{Methods}.
Average visibilities over $0.99 \pm 0.004$ were observed between 1540 and 1560~nm, and a sharp reduction was attained at lower wavelengths. This demonstrates that the device is capable of switching WDM signals operating in the described range.

\subsection*{Operation in optical networks}
\begin{figure}[b!]
  \centering
  \begin{subfigure}[b]{0.67\textwidth}
    \includegraphics[width=\textwidth]{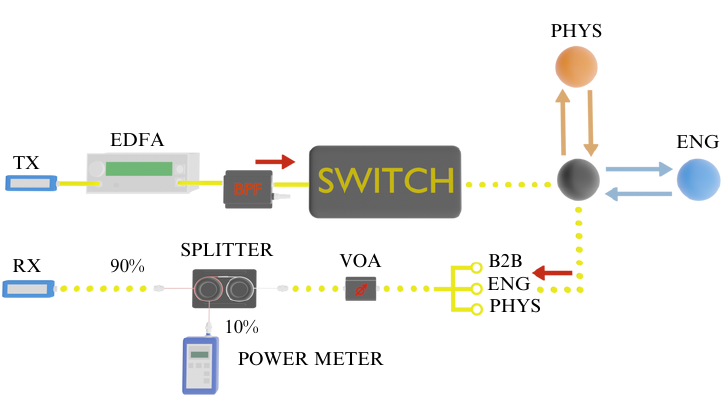}
    \caption{}
    \label{fig:SetupBER}
  \end{subfigure}%
  \begin{subfigure}[b]{0.33\textwidth}
    \includegraphics[width=\textwidth]
    {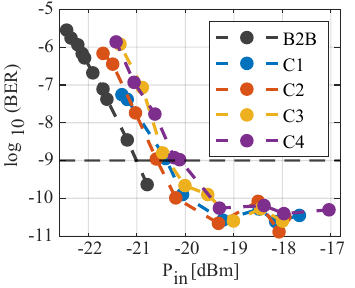} 
    \caption{}
    \label{fig:BER}
  \end{subfigure}
  \caption{\textbf{Transmission of 1 Gbps telecommunication signal.} \textbf{(a)}, Transmission setup. \textbf{(b)}, Received bit error rate as a function of the received power.}
  \label{fig:imagenes}
\end{figure}
An important feature of an optical switch is the potential to operate with conventional telecommunication signals. In this section we used a commercial small form-factor pluggable transceiver (SFP+) to generate and detect an optical signal to be switched in the device. The experimental setup shown in Fig. \ref{fig:SetupBER}. was used to evaluate the transceiver sensitivity in a back-to-back (B2B) configuration, as well as after transmission through the MCF switch connected to a field-installed MCF network.

The transceiver was driven by an independent FPGA with a pseudorandom binary sequence (PRBS) of length $2^{37}$ and was operated at 1 Gbps. Subsequently, the output was amplified by an EDFA and filtered by an optical band pass filter to remove out of band amplified spontaneous emission (ASE) noise. The amplified signal then entered the optical switch though core 1. The switch was operated to select one of the 4 output cores. After transmission, a variable optical attenuator (VOA) was used to control the optical power into the receiver and a 90-10\% splitter was used to monitor the received power. Finally, the FPGA calculated the bit-error-rate based on the received sequence. Note that to monitor the output of each core and to run the phase stabilization algorithm, the telecommunication signal was detected by the same photodiodes used in the previous section with 150 MHz of bandwidth. Additionally, a digital low pass filter was used on the sampling oscilloscope to average the fast transitions from the modulated signal, allowing the control algorithm to work at 0.8 MSps. 
Fig.~\ref{fig:BER} presents the measured bit error rate as a function of the received optical power. An average sensitivity penalty of 0.6 dB was observed at a BER of $10^{-9}$ after transmission through the switch, with a minimum penalty of 0.4 dB when selecting core 2 as output and a maximum penalty of 0.9 dB for core 4. This characterization shows the potential of the device to operate using conventional telecommunication signals. We note that to guarantee that the interference effect is present, the difference in the time of flight of the optical paths in the device needs to be smaller that the duration of the information carrying pulses.  

\begin{figure}[H]
  \centering
  \begin{subfigure}[b]{0.55\textwidth}
    \includegraphics[width=\textwidth]{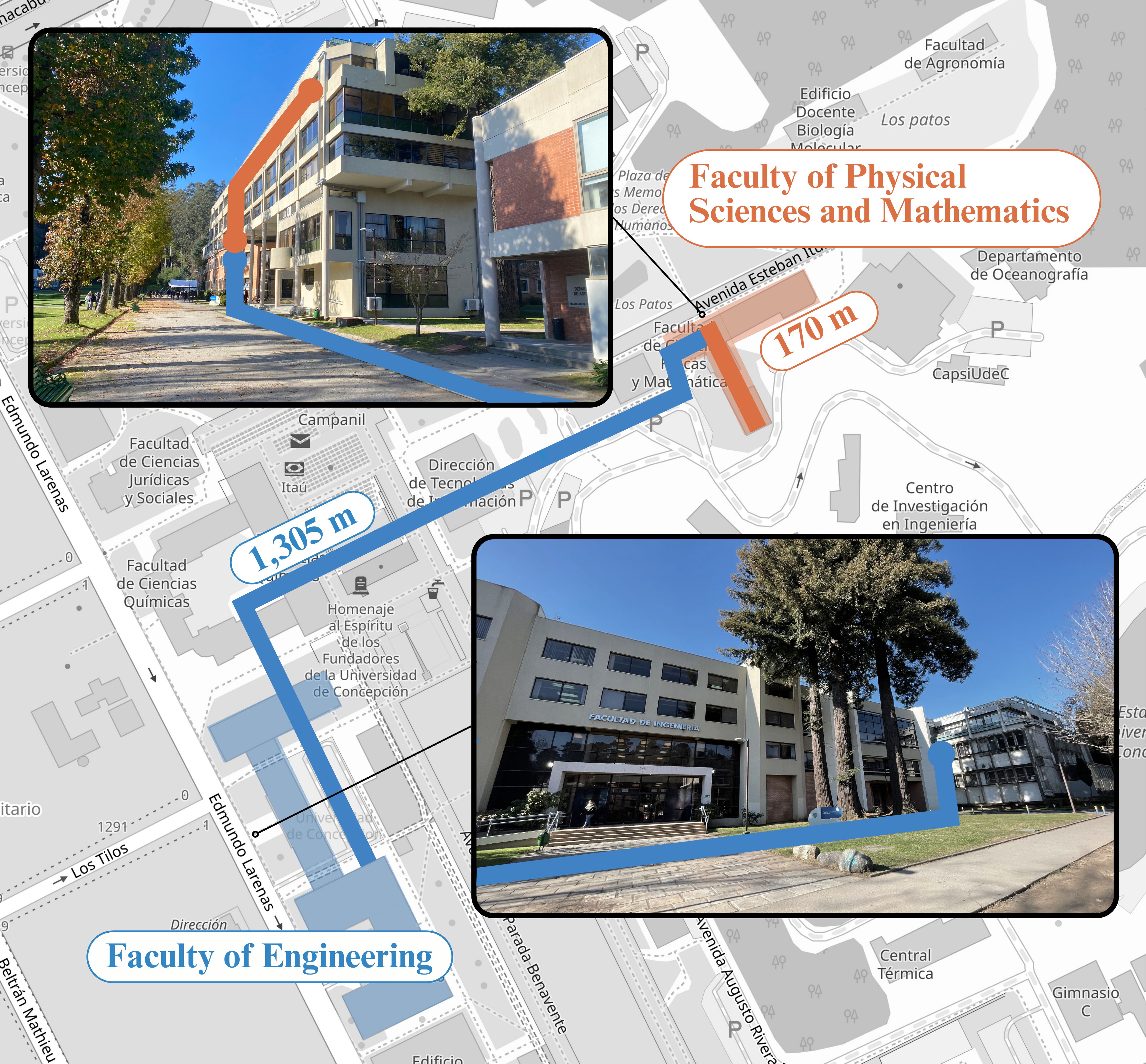}
    \caption{}
    \label{fig:Network Setup}
  \end{subfigure}%
  \begin{subfigure}[b]{0.40\textwidth}
    \includegraphics[width=\textwidth]{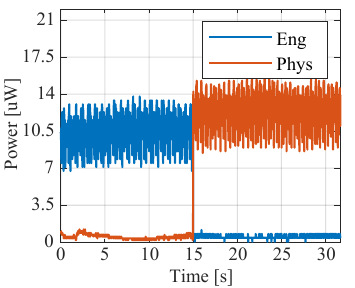}
    \caption{}
    \label{fig:imagen_switching}
  \end{subfigure}
  \caption{\textbf{Transmission of 1 Gbps telecommunication signal over MCF network.} \textbf{(a)}, Overview of the installed MCF links at Universidad de Concepción. \textbf{(b)}, Received optical power after propagation for both transmission links.}
  \label{fig:Network Results}
\end{figure}

Given the favorable characteristics shown by MCF switch in a table-top B2B scenario, what remained was to test the switch in a field-installed MCF network.
To do so, the switch was connected to a MCF network test-bed installed across the campus of the University of Concepción. The MCF network  connects the Optoelectronics laboratory located in the Faculty of Engineering with two Quantum Optics laboratories located across campus in the Faculty of Physical Sciences and Mathematics (FPSM) (see Fig.~\ref{fig:Network Setup}), and serves to evaluate transmission characteristics of MCF systems operating in a field environment. In this test, the transmitter was connected into transmission core 1 at the input of the MCF switch. The switch was located on the second floor of the FPSM and was used to select one of two possible transmission paths. To do this, the output of the switch was separated into 4 SMFs using a DMUX, subsequently, two outputs were connected to an internal MCF-link connecting the second and sixth floor of the FPSM with a total length of 170 m, and the other two were connected to an external MCF-link to the Faculty of Engineering with a total length of 1,305 m. Each link employed two MCFs, allowing the use of a loop-back configuration in order to detect the propagated signal using the SFP+ transceiver. At the receiver, the optical power was monitored by a pair of photodiodes.
Fig.~\ref{fig:imagen_switching} shows the received signal power at the output of the loopback configuration, with blue and orange representing the signal received from the path connected to the FPSM, respectively. Both signals show high-frequency intensity fluctuations corresponding to the pulses carrying the information, which are averaged due to the low bandwidth of the photodiodes.  
The device was used to perform switching successfully after 15 seconds of transmission. In both cases, error-free transmission was achieved and detected at the transceiver. 
In addition, IC-XT was measured after transmission. For the transmission link toward the Faculty of Engineering an IC-XT of -18 dB was obtained, while for the second link an IC-XT of -22dB was obtained. We note that the reported values match those reported in the previous section, indicating that no additional penalties were observed after transmission, other than the expected signal attenuation due to the longer transmission distances.

\section*{Conclusion /Discussion}  %
 
We have constructed and tested a multicore fiber core-selective switch, and demonstrated performance characteristics compatible with those of telecommunications networks. The switch is based on a multi-path interference device and uses GHz phase modulation for fast optical switching. A digital control system is used  to select the desired output core for signal routing, and also for phase stabilization.
The system successfully  routes signals between different cores of an MCF with a switching time of 0.7 $\mu$s, which is three orders of magnitude faster than the previous state-of-the art in MCF switch solutions. Additional performance characteristics were also favorable: an extinction ratio of 19 dB,  inter-core crosstalk between -12 and -22 dB per core, and an  average insertion loss of 7.7 dB for working periods of 2.5 $\mu$s and 30 $\mu$s. 
The switch was tested on telecommunications signals produced with a commercial SFP+ transceiver operating at 1 Gbps, in both a back-to-back laboratory setting as well as a field-installed MCF network, and achieved error-free transmission and low inter-core crosstalk in both scenarios.  To the best of our knowledge, this is the first demonstration of a high-speed optical switch for MCF architecture that is capable of operating in real-world field conditions.

\section*{Methods}

\subsection*{Core selective switch operation principle}

A multi-path Mach Zehnder (MZ) architecture can be used to build a MCF interferometer. We used a four-path MZ topology built using two four-core MCF-BS \cite{Carine:20}, in such a way that the first one divides an incoming signal and the second one recombines the signal at the end of the interferometer. The operation of a four-path interferometer based on MCF-BS can be described using:

\begin{equation}
\begin{pmatrix}
y_1 \\

y_2 \\

y_3 \\

y_4 \\
\end{pmatrix}
=
M_{BS_{4x4}}\cdot M_{\theta}\cdot M_{BS_{4x4}}\cdot
\begin{pmatrix}
x_1\\
x_2\\
x_3\\
x_4\\
\end{pmatrix},
\label{eq:4P-MZ}
\end{equation}

where $y_i$ and $x_i$ are the output and input optical fields of the MZ for the $i^{th}$ core respectively, $M_{BS_{4x4}}$ is a unitary matrix characteristic of a four-path MCF-BS \cite{Carine:20}, while $M_{\theta}$ is a unitary matrix, characteristic of the phase shift on the four optical paths given by:  
\begin{equation}
M_{\theta} = 
\begin{pmatrix}
e^{j\theta_1 } & 0 & 0 & 0 \\
0 & e^{j\theta_2} & 0 & 0 \\
0 & 0 & e^{j\theta_3} & 0 \\
0 & 0 & 0 & e^{j\theta_4} \\
\end{pmatrix}.
\label{EQ:Matrix_Interf}
\end{equation}
 Here, $\theta_i$ corresponds to the phase experienced by the signal when propagating through the \sw{$i^{th}$} path. In general, $\theta_i$ can be rewritten to consider phase fluctuation arising from external sources $\theta_{n,i}$ and a control phase applied by a phase modulator $\theta_{PM,i}$, yielding $\theta_i=\theta_{n,i}+\theta_{PM,i}$.
In Eq.\eqref{EQ:Matrix_Interf}, the insertion loss and polarization changes on each path are assumed to be equal.

Note that the optical power at the output of each core depends on the relative phase difference between every path of the interferometer. Considering that, if the phase fluctuations of each path $\theta_{n,i}$ are stabilized, the power at the output of the MCF-BS can be determined by a set of phases applied to each path using phase modulators $\theta_{PM,i}$.

The interferometer was evaluated by interfering light from only two paths between the MCF-BSs: the reference path $i=1$ and one of the others $i={2,3,4}$. The PM in the path under evaluation was driven by a symmetric triangular signal with peak-to-peak voltage of 10 V. Interference patterns at the output of each core and for each path combination are shown in \ref{fig:methods_2path_interference}. An average visibility of {$0.97 \pm 0.013$ was observed for all the studied outputs and path combinations. 
Subsequently, with the phase fluctuations of the intereferometer the interference of the 4 paths was evaluated, and is shown in Fig.~\ref{fig:methods_4x4}. In this scenario an average visibility of 0.98 was obtained.
Finally, the interference visibility was evaluated across the C-band using a tuneable laser showing a quadratic tendency, shown in Fig.~\ref{fig:methods_Cband}. Here the values reported in section Core Switching are presented. 
The high visibility of the interference curves shows that the proposed device is suitable for switching operation.

\begin{figure}[H]
    \centering
    \begin{subfigure}[b]{0.33\textwidth}
        \includegraphics[width=\textwidth]{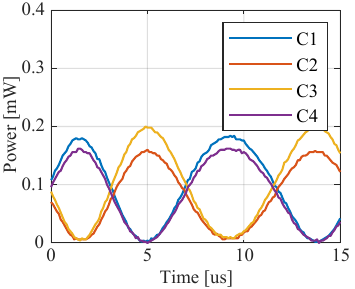}
        \caption{}
        \label{fig:methods_2path_interference}
    \end{subfigure}
    \hfill
    \begin{subfigure}[b]{0.33\textwidth}
        \includegraphics[width=\textwidth]{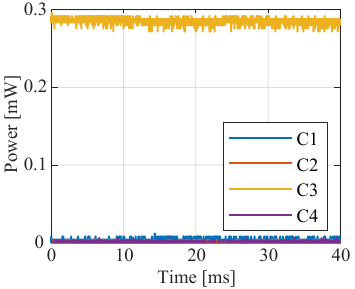}
        \caption{}
        \label{fig:methods_4x4}
    \end{subfigure}
    \hfill
    \begin{subfigure}[b]{0.33\textwidth}
        \includegraphics[width=\textwidth]{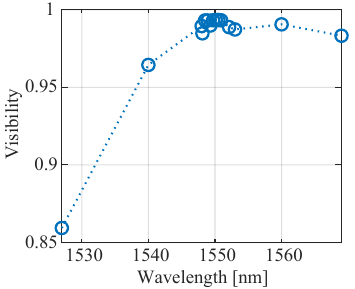}
        \caption{}
        \label{fig:methods_Cband}
    \end{subfigure}
    \caption{\textbf{Characterization of interference in a four-path Mach Zehnder interferometer.} \textbf{(a)}, Interference patterns between two paths. \textbf{(b)}, Stabilized interference between all 4 paths. \textbf{(c)}, Visibility across the C-band.}
    \label{fig:Path_combinations}
\end{figure}

\subsection*{Active phase stabilization}
As the output of the device depends on the phase relationship between the light in each propagation path, phase fluctuations due to environmental factors lead to excess insertion losses and inter-core cross-talk. In addition, unwanted phase fluctuations will continuously couple optical power from one core to another. 
A digital control system was used to mitigate the influence of phase fluctuations in the device.
The digital system introduces a set of control phases $\phi_{c,i}$, that were obtained using the nonlinear process known as perturb and observe \cite{Alik2016}, where phase values are searched by sequential variations of the phase in each path so that the intensity is maximized at the output of a single core. 

\bibliography{Biblio}

\section*{Acknowledgements}

This work was supported by Fondo Nacional de Desarrollo Científico y Tecnológico (ANID) (Grants No.  1200266, 1240746, 1240843, 1231826, 1220960, 1231940), ANID – Millennium Science Initiative Program – ICN17\_012, ANID SUBDIRECCIÓN DE INVESTIGACIÓN APLICADA  Folio ID22I10262 and ANID Basal Fund FB0008. The authors acknowledge OpenStreetMap\cite{OpenStreetMap} for providing the map of Concepción City, Chile.

\section*{Author contributions statement}

G.S., J.C, G.L and S.P.W conceived the experiments,  C.M.,M.R.F. and D.A. conducted the experiments, All authors analyzed the results and reviewed the manuscript. 




\end{document}